# EVALUATION AND MITIGATION OF TRAFFIC RELATED CONCERNS IN KATHMANDU VALLEY

Aadarsha Bhusal*, Ankit Jha, Anish Khanal, Aayush Awasthi, Aman Kumar Mahato
*St. Xavier's College, Maitighar, Kathmandu, Nepal*

**ABSTRACT**
*Motor cars are more prevalent on the highways of major cities like Kathmandu as a result of the raging increase in urban population. As a result, the residents of this city are accustomed to experiencing frequent traffic jams. The fact that every person is forced to spend an average of 120 hours a year doing nothing productive due to traffic congestion makes this very clear. Evidently, identifying the problem does not always result in a fix. However, creating intelligent traffic management systems with intelligent traffic signals and early warning systems could be a successful substitute. Furthermore, creative road engineering, as demonstrated by features like road diet and roundabouts, is a popular treatment. However, the primary issue is the preference for private vehicles over public transportation. It can definitely help to reduce these worries to cultivate a society-friendly mentality along with effective, inventive approaches.*

KEYWORDS: Congestion, Public transportation, Intelligent traffic lights, ETS, Road diet, and Roundabouts

## 1. INTRODUCTION

The Kathmandu city is home to a massive fleet of more than 100,000 cars. Large highways alone won't be able to stop this chaos because more vehicles only lead to more congestion. This analysis demonstrates that the usage of private vehicles is the main issue, helped along by the faces and urban legends of the general population. Other difficulties include disobeying driving laws, poor road quality, and traffic lights. In addition, this research considers other less well-known issues such poorly managed intersections, broken streetlights, and recklessness on the part of locals.

It has never been a solution to make highways bigger. The relationship between road width and traffic congestion shows a straight line, i.e., congestion rises linearly with the width of the road. Therefore, as further explored in this paper, the solution to this problem calls for unconventional approaches as well as social indulgence. Congested traffic can be helped to flow by the deployment of technological innovations like road diets and enhanced intersections like roundabouts.

In a similar vein, this paper recognizes the advantages of smart traffic management, which employs sensors, the cloud, and integrated traffic signals to reduce traffic congestion, as is seen in northern countries. Making new rules serves no purpose as long as there are those who are unable to abide by them. Public awareness will always be the most crucial factor in this aspect. While science cannot alter your perspective on the world, it may unquestionably alter what you observe. Even if technical advancements might not be the whole answer, they can be extremely important in the context of Kathmandu if used correctly and on schedule. The social component, when combined with it, can undoubtedly raise the bar for traffic control in this city.

In this essay, we assess the current traffic management situation in Kathmandu, list the issues, and offer some tried-and-true as well as emerging solutions to address these issues.

## 2. LITERATURE REVIEW

Some of the main reasons of congestion are rapid population growth, increased urbanization, bad public transportation systems, inadequate/unplanned transportation infrastructure, and an increase in the number of personnel vehicles (Kumar and Das, Study on road traffic congestion: A Review, 2021).

According to the same study's findings, key factors in reducing congestion include better traffic management and control, better public transportation, increased funding for transportation infrastructure, the use of cutting-edge technology, and overall coordination of land-use and transportation policies.

The Internet of Things (IOT), cloud computing, 5G, and big data are just a few of the cutting-edge technologies that are being used in a report by Humayun, et al. that was published in the Journal of Advanced Transportation to help conventional traffic management systems and effectively address the stated issue.

In the field of smart cities, intelligent data flow management has garnered a lot of interest. Making the best judgments in the shortest period of time requires learning how to use the vast amounts of data that smart city infrastructure and technology collect (Multi-Agent System for Intelligent Urban Traffic Management Using Wireless Sensor Networks Data,Muntean, 2021). According to a selected reference state, the condition of traffic flow on a transportation facility that is congested is one that has high densities and slow speeds (with low densities and high speed). (Timalsena, Marsani, and Tiwari, 2017). Impact of Traffic Bottleneck on Urban Road: A Case Study of Maitighar-Tinkune Road Section The impact of the congestion on the Maitighar-Baneshwor-Tinkune road stretch is examined in the same paper. In addition, one of the study's primary goals was to ascertain the delay caused by the bottleneck in this section of the road in order to calculate the total loss in human capital hour due to congestion during peak hours.

### 3. OBJECTIVES

The achievement of the following goals is the primary focus of this report:
- To assess the Valley of Kathmandu's overall traffic flow
- To determine the typical delay caused by traffic congestion
- To identify the population segment using different public and private means of transportation.
- To identify various traffic offenses and abnormalities that are obstructing the flow of traffic.
- To Recognize the role of the public and government in traffic management
- To find unconventional methods for road design that would help traffic flow
- To evaluate the suitability and efficiency of such actions in the valley setting.
- To represent a practical conclusion on the assessment and easing of traffic-related issues in the valley.

### 4. METHODOLOGY

#### 4.1 Example of Study Subject

In the Kathmandu Valley, the general populace—of whom students made up a sizable portion—was the sample population. Due to traffic congestion, this region highlighted the plight and issues of the general populace. It also demonstrated how widely used various forms of transportation are.

#### 4.2 Resources and Methods

A generic public questionnaire and a special interview created for the traffic spokesperson were the easier data collection options. Additionally, developing solutions required the use of simulation programs like Cities Skylines, Tinkercad, and video editing tools.

#### 4.3 Information sources

The survey, the first-hand accounts, and the interview with a traffic spokesperson are the main information sources. This information shows how frequently accidents occur as well as how many different types of transportation are used. The secondary sources include a wide range of websites, a small number of publications, and research papers that helped us learn about some crucial data and solutions.

### 5. FINDINGS

#### 5.1 The narrative of numbers

The statistics provide a story that is consistent with what is immediately seen. The provided piechart demonstrates the widespread preference for private vehicles, particularly two-wheelers, over public ones.

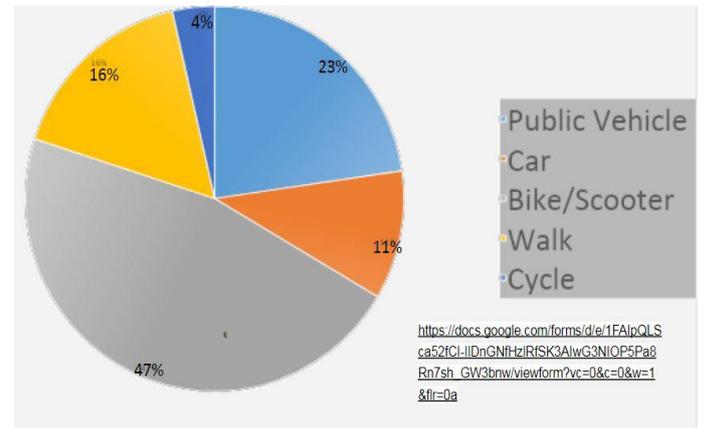

According to study data, a person spends an average of 15-20 minutes per day in traffic, which adds up to a whopping 120 hours per year. By doing this, the year is cut by five absolutely useless days.

### ACCIDENT DATA

| क्र.सं. | महिना | दुर्घटना संख्या | मृत्यु | गम्भिर घाइते | सामान्य घाइते |
|---|---|---|---|---|---|
| 1. | श्रावन | ९२८ | १२ | २४ | ७८७ |
| 2. | भाद्र | ८८२ | ८ | २२ | ७७८ |
| 3. | असोज | ७४२ | ८९ | ८९ | ५९७ |
| 4. | कार्तिक | ६२८ | १६ | २१ | ४८४ |
| 5 | मंसिर | ७६० | १३ | १८ | ४८० |
| 6. | पुष | ८४९ | ११ | १४ | ६७६ |

The graph above demonstrates how citywide traffic issues and accidents change over the course of the year. The months of Ashwin and Kartik have the fewest accidents, which clearly reflects the relatively new beginnings of the traffic flow. This is primarily caused by the shutdown of offices and schools. The graph below, which considers the alarming trend of rising vehicle population in the valley, reveals further noteworthy facts.

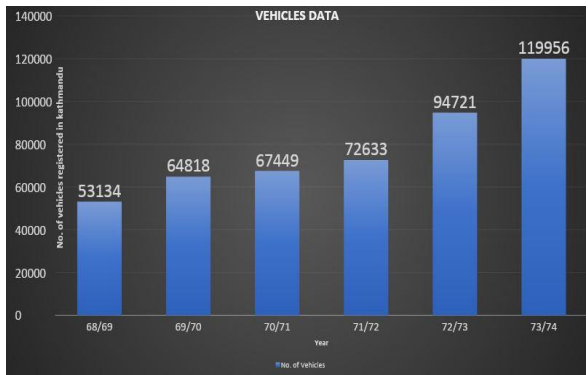

As speculated by the local governing bodies, the numbers have alarmingly increased in the recent couple of years despite the effects of the COVID-19 pandemic.

### 5.2 Wheels VS Roads

The ever growing population of automobiles in the Kathmandu valley poses some grave concerns, particular of which are discussed in the lines to follow:

- **Disproportionate increase in the number of private vehicles:** The number of vehicles, in particular private vehicles, is shooting out in alarming rate. The poor transport planning is responsible for traffic congestion. The numbers of vehicles especially the two wheelers are increasing due to the lack of appropriate measures to tackle this unprecedented rise in the ownership of private vehicles. Middle and upper middle class people prefer to drive their own vehicles for lack of public confidence in public transport, erratic behaviour of the drivers, overcrowding and long stops at various places.

- **The stagnation of Road Network:** The existing road network of the valley simply cannot cope with the rising population and ownership of private vehicles. The lack of proper planning and implementation for increasing the road network and lack of proper maintenance of the existing roads is seriously contributing to valley's woes.

- **Haphazard State of Vehicle Parking System and street vendors:** The state of Vehicle parking especially in the core areas is extremely haphazard – beyond the level of comprehension for people who walk. Even where you have to pay nominal parking fee, the vehicles can be seen being parked in the footpath or the parking stretches up to the middle of the main road. This problem is just magnified by the presence of unmanaged street vendors – which just gives you the feeling that you can just open the shop wherever you want.

- **Recklessness of People:** One of the major reasons for the traffic issue of Kathmandu is the recklessness of people. Many people are seeming not to be following traffic rules which automatically invites various problems. Activities like crossing the lane, not following the traffic light, taking the vehicles haphazardly in crowded places (people take car inside the narrow road of Asan), drinking and driving, etc. increase the chances of congestion and accidents.

### 5.3 The Eureka Zone- Solutions

Problem shall be inevitable but the search for solutions is what keeps the spirit of human society alive. In such periphery, the below mentioned solutions can be of paramount importance:

### 5.3.1 Simulation of smart traffic light management system

This system automatically computes the optimal green signal time based on the current traffic at the signal. This will ensure that the direction with more traffic is allotted a green signal for longer duration of time as compared to the direction with lesser traffic.

The main objective of this system is to design a traffic light controller based on computer vision that can adapt to the current traffic situation. This system aims to use live video feed from the CCTV cameras at traffic junctions for real-time traffic density calculation by detecting the vehicles at the signal and set green signal time accordingly. The vehicles are classified as car, bike, bus/truck, or rickshaw to obtain a more accurate estimate of the green signal time.

*Advantages*

- Real time light switching according to current traffic density
- Virtually no new hardware to be installed
- Less expensive than sensors
- Autonomous: no need of manpower

*Proposed system model*

- The system will pass a snapshot from the CCTV cameras at traffic junctions for real time traffic density calculation using Image Processing and Computer Vision.
- The scheduling algorithm will use this traffic density and appropriately set the optimal green signal time for each signal, and update the red signal times of the other signals.

*Factors considered in signal switching algorithm:*

- The processing time of the algorithm to calculate traffic density
- The number of lanes
- Lag each vehicle suffers during start-up
- The nonlinear increase in lag suffered by vehicles which are at the back
- The maximum and minimum green signal time that can be set (to prevent starvation of the lane with less traffic)

*Flow chart:*

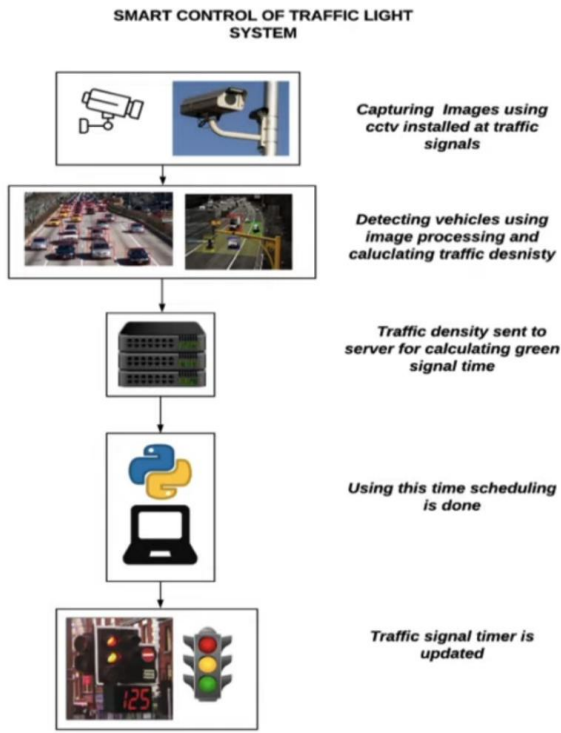

*Result:*

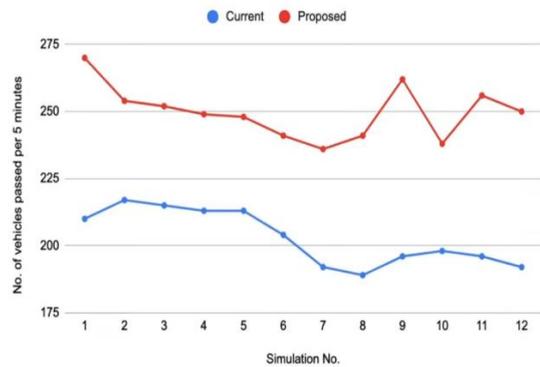

As we can see, with all conditions alike, the adaptive system was able to pass **2997** vehicles while the current static system could only pass only **2435** vehicles in **1 hour,** which means **562** more vehicles. Thus, the proposed adaptive system improves the performance by over 23%. The adaptive system on an average allows **48** more vehicles to pass every **5 minutes** as compared to the static system.

### 5.3.2 Reversible lane system

A section of Araniko highway between Jadibuti and Koteshwor intersection has been chosen as a case study for evaluation of reversible lane application. The section connects the two major cities of the Kathmandu valley: Bhaktapur and Kathmandu.

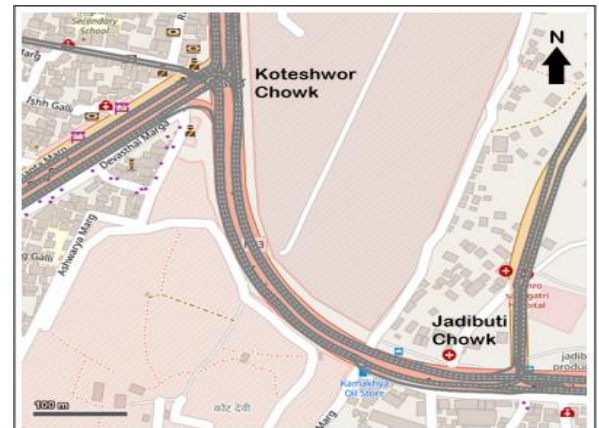

Figure shows the location map and layout of Jadibuti-Koteshwor section. The section connects two intersections at Jadibuti and Koteshwor. As shown in the figure, the Jadibuti intersection has major approaches to and from Lokanthali, Pepsicola and Koteshwor while the Koteshwor intersection has major approaches to and from Jadibuti, Balkumari and Tinkune.

| Current System | | | | | | Proposed Adaptive System | | | | | |
|---|---|---|---|---|---|---|---|---|---|---|---|
| Simulation No. | Lane 1 | Lane 2 | Lane 3 | Lane 4 | Total | Simulation No. | Lane 1 | Lane 2 | Lane 3 | Lane 4 | Total |
| 1 | 67 | 74 | 51 | 18 | 210 | 1 | 111 | 86 | 42 | 31 | 270 |
| 2 | 78 | 73 | 47 | 19 | 217 | 2 | 105 | 83 | 38 | 28 | 254 |
| 3 | 80 | 73 | 33 | 29 | 215 | 3 | 100 | 96 | 36 | 20 | 252 |
| 4 | 76 | 71 | 39 | 27 | 213 | 4 | 96 | 75 | 56 | 22 | 249 |
| 5 | 77 | 66 | 44 | 26 | 213 | 5 | 93 | 89 | 42 | 24 | 248 |
| 6 | 74 | 72 | 37 | 21 | 204 | 6 | 77 | 97 | 37 | 30 | 241 |
| 7 | 65 | 73 | 36 | 18 | 192 | 7 | 76 | 82 | 48 | 30 | 236 |
| 8 | 60 | 68 | 33 | 28 | 189 | 8 | 71 | 92 | 48 | 30 | 241 |
| 9 | 49 | 83 | 36 | 28 | 196 | 9 | 85 | 98 | 48 | 31 | 262 |
| 10 | 57 | 70 | 46 | 25 | 198 | 10 | 79 | 92 | 37 | 30 | 238 |
| 11 | 53 | 70 | 39 | 34 | 196 | 11 | 110 | 105 | 24 | 17 | 256 |
| 12 | 55 | 70 | 29 | 38 | 192 | 12 | 76 | 87 | 43 | 44 | 250 |
| | | | | | 2435 | | | | | | 2997 |

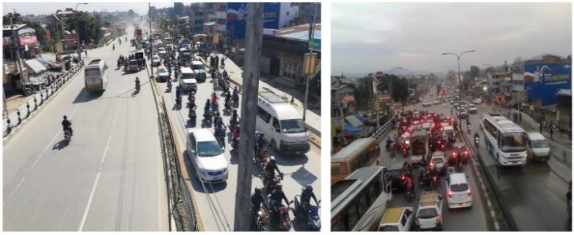

Variation of direction flow during morning and peak hour

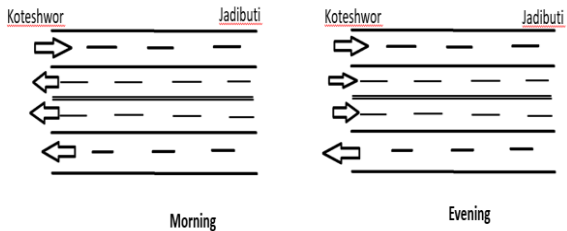

Proposed reversible plan

For this reversible lane model, an eight lane highway is taken. Normally, 4 lanes are used by vehicles approaching from Jadibuti and the same for vehicles from Koteshwor. Currently, this system is ineffective in decreasing the traffic congestion in this area.

In the proposed system, the central 4 lanes are reversed according to the traffic density. However, the lanes towards the outer side are not changed at all. As seen before, the traffic from Jadibuti to Koteshwor is maximum in the morning peak hours. Therefore, the 4 lanes in the center are directed towards Koteshwor from Jadibuti. But, the traffic from Koteshwor to Jadibuti is maximum in the evening peak hours. Therefore, the 4 lanes in the center are directed towards Jadibuti from Koteshwor, just opposite of the morning period.

As per the results obtained from simulation conducted in VISSIM software, the reversible lane plan can significantly reduce the queue lengths at the junctions specially that of the through traffic in the road section in both morning and evening peak hours. Queue lengths at the most queued locations near the junctions, Lokanthali in the morning can be reduced by 50%, and Tinkune in the evening can be reduced by 54%. Similarly, improvement in junction traffic flows in terms of reductions in queue lengths 24% in the morning and 35% in the evening can be achieved at Pepsicola and Balkumari, respectively.

Table below shows a comparison of travel time in seconds between the intersections in morning and evening peak hours for the current and reversible model. The results show that the reversible lane plan improves the travel time of the section by 11% on average. RLS of morning peak hour particularly reduces the travel time of the section in both direction, 19% from Jadibuti to Koteshwor and 8% from Koteshwor to Jadibuti. Reversible lane plan at the evening peak hour also reduces the travel time of the section in direction from Jadibuti to Koteshwor by 21% but looks indifferent to the travel time in opposite direction. Therefore, reversible lane plan at the morning peak hours is more effective compared to that at the evening peak hours.

| Travel time (sec) | | J-K | K-J |
|---|---|---|---|
| Morning | Current | 195 | 140 |
| | Reversed | 157 | 129 |
| Evening | Current | 169 | 82 |
| | Reversed | 134 | 86 |

Comparison of traveled time in current and reversed model

### 5.3.3 Lane Discipline in Traffic Management

Lane discipline refers to the responsible and safe usage of lanes on a road by drivers. Proper lane discipline is crucial in maintaining traffic flow and reducing the risk of accidents.

Here are some key principles of lane discipline in traffic management:

1. Stay in your lane: Drivers should always drive in the lane that corresponds to the direction they are traveling. Changing lanes frequently or without signaling can cause confusion and increase the risk of accidents.
2. Use signals: Drivers should use their turn signals to indicate when they are changing lanes or turning. This gives other drivers an opportunity to anticipate the change and adjust their own driving accordingly.
3. Avoid tailgating: Tailgating, or following too closely behind another vehicle, increases the risk of accidents. Drivers should maintain a safe following distance to allow time to react to changes in traffic flow.
4. Keep right except to pass: In most countries, slower vehicles are expected to drive in the right-hand lane, while faster vehicles use the left lane for passing. This helps to maintain a smooth flow of traffic.
5. Avoid blocking the passing lane: Drivers should avoid lingering in the passing lane if they are traveling slower than the flow of traffic. This can cause congestion and frustration for other drivers.

### 5.3.4 Localization of institutions

Restriction of schools, colleges, and institutions in a certain area can help in managing traffic by reducing the number of vehicles on the road during peak hours, when traffic is often the most congested. By restricting the number of schools, colleges, and institutions in an area, there will be fewer students and employees traveling to and from these institutions, which can help reduce traffic congestion and improve traffic flow.

Some of the specific ways that restriction of schools, colleges, and institutions can help in managing traffic are:

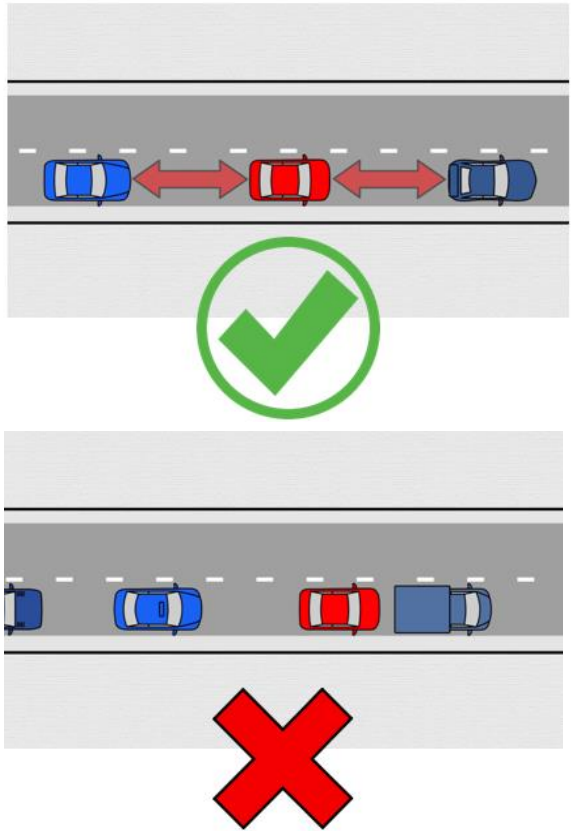

1. Reduced peak-hour traffic: By limiting the number of institutions in a certain area, the number of vehicles on the road during peak hours can be reduced. This can help improve traffic flow and reduce congestion during these times.
2. Improved safety: Limiting the number of institutions in a certain area can reduce the number of vehicles on the road during peak hours, which can improve safety for all road users, especially pedestrians and cyclists.
3. Increased public transportation usage: By reducing the number of vehicles on the road, restriction of schools, colleges, and institutions can encourage people to use public transportation instead. This can help reduce traffic congestion and improve the efficiency of public transportation systems.

In conclusion, restriction of schools, colleges, and institutions in a certain area can help in managing traffic by reducing the number of vehicles on the road during peak hours, which can improve traffic flow, safety, air quality, and public transportation usage. However, it's important to carefully consider the potential impacts of such restrictions and to engage with relevant stakeholders, such as students, employees, and institutions, to ensure that the restrictions are implemented in a way that is equitable and sustainable.

### 5.3.5 Improved Junctions
*Roundabouts:*

The use of roundabouts has accounted for reduction in vehicular fatalities as well as improvement in traffic flow over the years. In accordance with this fact, most of the European Countries use roundabouts as their go to junctions. Another interesting feature is that it doesn't require traffic lights to operate.

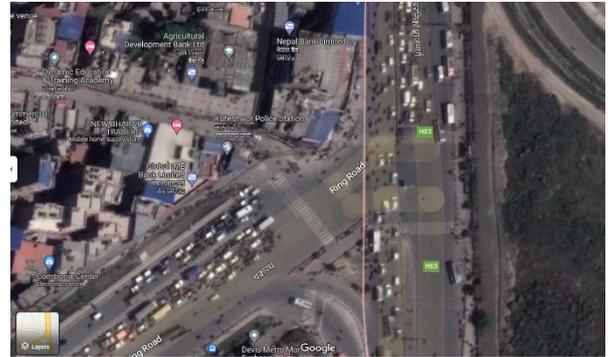

Normal junction at Koteshwor

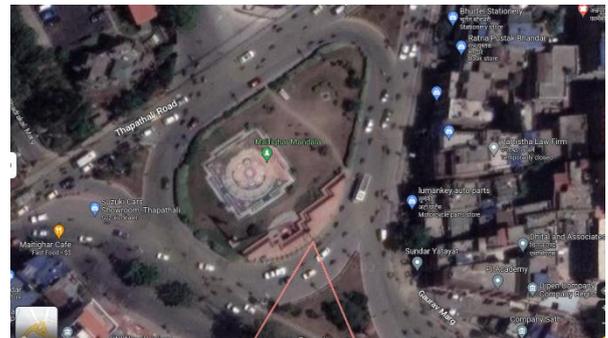

Roundabout at Maitighar

### 5.3.6 Time Shift
*Marginalizing Traffic Flow*

One of the major reasons behind frequent traffic jams is the coincidental timings of academic and administrative institutions. Accordingly, shifting the active hours of these two domains can surely benefit the traffic flow. As revealed by the recent notice regarding such traffic shift in Chandigarh, the traffic congestion decreases by 10% on an average.

### 6.CONCLUSION

The Kathmandu valley has always struggled with traffic congestion, which has geometrically gotten worse as human density has increased. We lose both time and life due to traffic problems, thus finding pertinent and practical solutions has become necessary.

In addition to engineering solutions, the use of social innovations such as changing the working hours of service industries, using ENS, and discouraging the use of two-wheelers while promoting the use of public vehicles with appropriate governmental or municipal regulation can effectively address traffic issues. Similar to how building roundabouts at busy intersections improves traffic flow and gets rid of backups. In addition, maintaining lane discipline and using the reverse lane system are efficient ways to deal with traffic that is already flowing.

### 7. FUTURE SCOPE

This in-depth investigation is solely based on the most recent study of the Kathmandu valley's traffic congestion situation. In order to smoothly regulate traffic, the paper examines and analyzes the relevance and usefulness of every feasible effective measure. The methods have been effectively implemented in numerous places throughout the world, and they are supported by a variety of published statistics and publications from reliable sources.

As a result, this paper is a great resource for learning how to solve traffic problems and can be used as a guide for similar research in the future. Additionally, this can be used as a reference when undertaking research of a similar nature on several growing heavily populated cities, such as Kathmandu.

### 8. LIMITATIONS OF THE STUDY

Although the research primarily focuses on the Kathmandu valley's traffic mitigation measures, the references cited for those methods might not be appropriate for a developing nation like Nepal.

The majority of the remedies also center on getting people to change their attitudes regarding traffic enforcement, which is a huge subject in and of itself. Similar to how inadequate research on traffic issues in the Kathmandu valley made this research more difficult.